\documentclass[aps,prb,twocolumn,epsfig,showkeys,superscriptaddress]{revtex4-1}
\usepackage{graphicx}%Include graphics file
\usepackage{color}
\usepackage{bm}% bold math
\usepackage{csquotes}
\makeatletter

\newcommand{\Rmnum}[1]{\expandafter\@slowromancap\romannumeral #1@}
\makeatother

\begin{document}
\title{\bf Huge magnetoresistance and ultra-sharp metamagnetic transition in polycrystalline
\boldsymbol{${Sm_{0.5}Ca_{0.25}Sr_{0.25}MnO_3}$}}
\author{Sanjib Banik}
\affiliation{CMP Division, Saha Institute of Nuclear Physics, HBNI, 1/AF-Bidhannagar, Kolkata 700 064, India}
\author{Kalipada Das}
\affiliation{CMP Division, Saha Institute of Nuclear Physics, HBNI, 1/AF-Bidhannagar, Kolkata 700 064, India}
\affiliation{Department of Physics, Seth Anandram Jaipuria College, 10-Raja Naba Krishna Street, Kolkata-700005, India}
\author{Tapas Paramanik}
\affiliation{CMP Division, Saha Institute of Nuclear Physics, HBNI, 1/AF-Bidhannagar, Kolkata 700 064, India}
\affiliation{Indian Institute of Technology, Kharagpur- 721302, India}
\author{N. P. Lalla}
\affiliation{UGC-DAE Consortium for Scientific Research, University Campus, Khandwa Road, Indore-452017, India}
\author{Biswarup Satpati}
\affiliation{Surface Physics and Material Science Division, Saha Institute of Nuclear Physics, HBNI, 1/AF-Bidhannagar, Kolkata 700 064, India}
\author{Kalpataru Pradhan}
\email{kalpataru.pradhan@saha.ac.in}
\affiliation{CMP Division, Saha Institute of Nuclear Physics, HBNI, 1/AF-Bidhannagar, Kolkata 700 064, India}
\author{I. Das}
\email{indranil.das@saha.ac.in}
\affiliation{CMP Division, Saha Institute of Nuclear Physics, HBNI, 1/AF-Bidhannagar, Kolkata 700 064, India}
\email{indranil.das@saha.ac.in}
%-----------------------------------------------------------------------------

\begin{abstract}
Large magnetoresistive materials are of immense interest for a number of spintronic
applications by developing high density magnetic memory devices, magnetic
sensors and magnetic switches. Colossal magnetoresistance, for which resistivity
changes several order of magnitude (${\sim10^4 \%}$) in an external magnetic
field, occurs mainly in phase separated oxide materials, namely manganites, due
to the phase competition between the ferromagnetic metallic and the
antiferromagnetic insulating regions. Can one further enhance the
magnetoresistance by tuning the volume fraction of the two phases? In this work,
we report a huge colossal magnetoresistance along with the ultra-sharp
metamagnetic transition in half doped ${Sm_{0.5}Ca_{0.25}Sr_{0.25}MnO_3}$
manganite compound by suitably tuning the volume fraction of the competing phases.
The obtained magnetoresistance value at 10 K is as large as $\sim10^{13}\%$ in a
30 kOe external magnetic field and $\sim10^{15}\%$ in 90 kOe external magnetic field
and is several orders of magnitude higher than any other observed magnetoresistance
value reported so far. Using model Hamiltonian calculations we have shown that
the inhomogeneous disorder, deduced from tunneling electron microscopy,
suppresses the CE-type phase and seeds the ferromagnetic metal in an
external magnetic field.
\end{abstract}

%\pacs{ 75.47.Lx, 73.63.Bd} \maketitle
%\keywords{Manganites, Magnetoresistance, Disorder}
\maketitle
%------------------------------------------------------------------------------
\section*{Introduction}
%------------------------------------------------------------------------------

For the last several years  search for  materials  with large magnetoresistance
(MR) and the studies on related  phenomena are on the forefront of the worldwide
research activity~\cite{Moritomo,Shimakawa,Uehara,Siwach,Ali,Tafti,Baldini}
due to its widespread application in the field of magnetic sensor, magnetic memory
devices, magnetic switches etc~\cite{asamitsu,bibes,sawa,rubi,hoffman}. The highest
value of MR in principle can be achieved if the resistivity value of the material
can be transformed from an extreme insulating material (like mica) to a very good
metallic one (like Copper) by applying magnetic field.

%***************************************************************************************
\begin{figure*}
\includegraphics[width=0.95\textwidth]{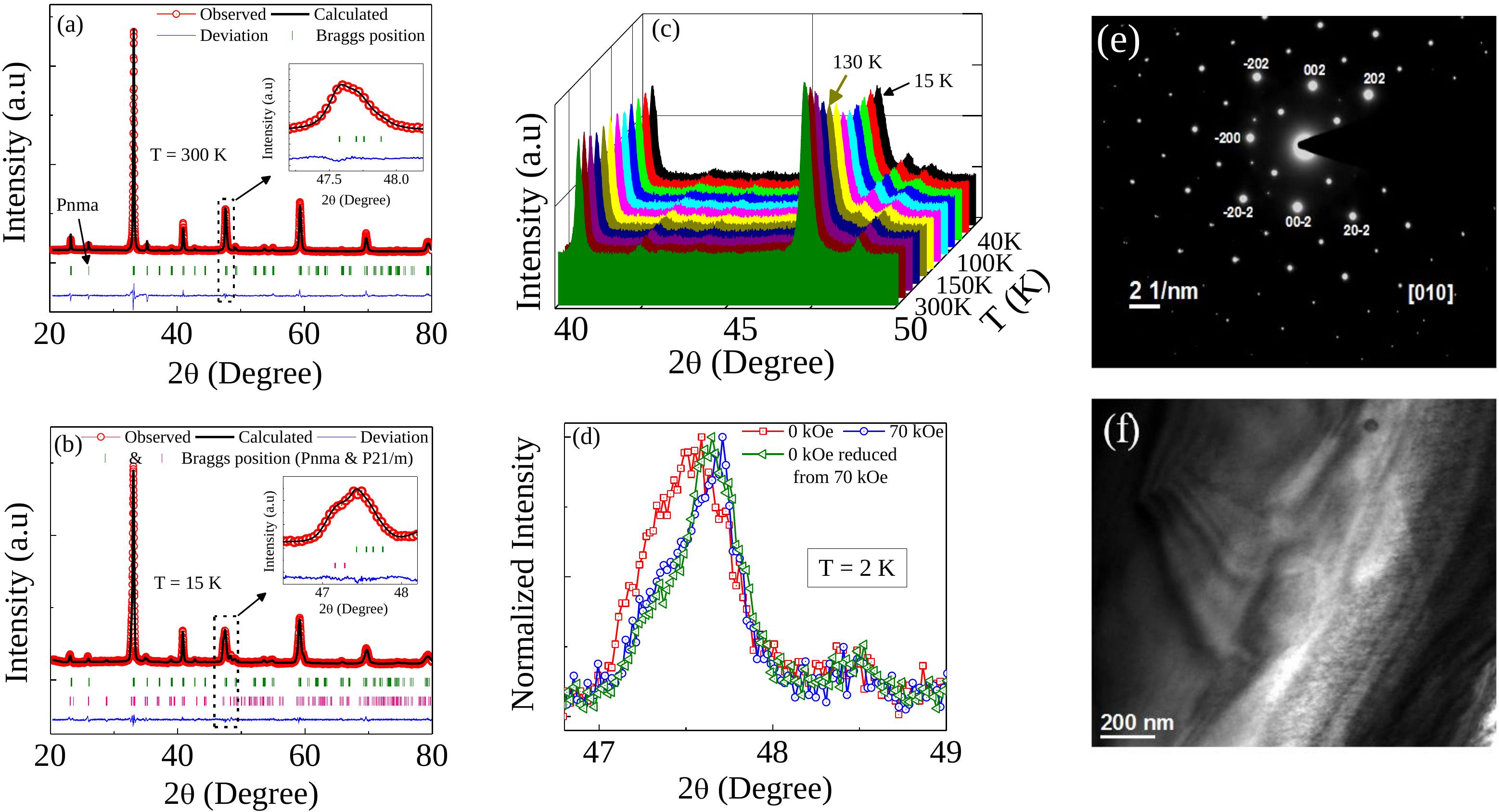}
\centering \caption[ ]
{\label{TEM} XRD and TEM analysis: Panel (a) and (b) shows the profile fitting of the room
temperature (RT) and low temperature (15 K) XRD data using Pnma and (Pnma + P21/m)
space group. Inset of (a) and (b) shows the fitting of the peaks at $2\theta =47.5^0$
where new monoclinic phase (P21/m) appears at low temperature. Panel (c) displays the
evolution of the appearance of the new peak of P21/m space group with temperatures.
Panel (d) shows the XRD line width modification in the presence of zero magnetic field,
70 kOe magnetic field as well as after removing the field. Panel (e) displays a typical
[100] zone axis ED patterns at room temperature. The pattern was indexed using
orthorhombic structure (Pnma). Panel (f) demonstrates [001] bright-field
image recorded at 100 K from TEM measurements. For HRTEM analysis please see
Supplementary section I(B).
}
\end{figure*}
%***************************************************************************************

In perovskite manganite, insulating state is observed in charge ordered antiferromagnetic
(CO-AFM) sample which generally appears near the half doping. Critical magnetic field,
that is required to melt the charge-ordered antiferromagnetic state, increases with
decreasing the bandwidth. CO-AFM state can be weakened by introducing ferromagnetic
proximity like: (i) by effectively increasing the bandwidth of the ${e_g}$ electrons
via substituting larger cations at the A-sites~\cite{Arulraj,TokuraCMR,Nigam,Mavani},
(ii) by B-site doping (e.g, Cr, Ru doping on Mn sites)~\cite{Martin,Hervieu1,pradhan1},
(iii) by making FM-AFM core-shell nanostructures (or nanoparticles)~\cite{Das,Das2,Dong}.
Thus by inducing the ferromagnetic phase fraction in an antiferromagnetic manganite or
in other word engineering electronic phase separation is an effective tool to enhance
the magnetoresistance. Recently, great amount of effort has been devoted to control
the electronic phase separation in bulk and low dimensional
manganites~\cite{Zhu,Chen,Chai,Shao,Zhang,elovaara1,ZhangNL}.

It is known that CO-AFM state in a low bandwidth ${Sm_{0.5}Ca_{0.5}MnO_3}$
(SCMO) is very stable\cite{Tokura2} and ${\sim500}$ kOe [at temperature (T) = 4.2 K]
critical magnetic
field ($H_{CR}$) is required to destabilize the CO state. On the other hand, CO-AFM
and FM metallic phases coexist in ${Sm_{0.5}Sr_{0.5}MnO_3}$ and is metallic at low
temperatures\cite{Tomioka}.
In our case we prepare ${Sm_{0.5}Ca_{0.25}Sr_{0.25}MnO_3}$ [SCSMO] by replacing half of
the $Ca^{2+}$ ions by $Sr^{2+}$ ions. This substitution reduces the charge-ordering
temperature ($T_{CO}$) and Neel temperature ($T_N$) of SCSMO by $\sim$50 K
from its parent compound SCMO\cite{TokuraCMR}. Although, SCSMO remains a strong insulator
[$\rho$ $\sim$ 10$^{12}$ Ohm-cm] at low temperature,
but the critical field decreases to ${~48}$ kOe (for SCSMO) from ${~500}$ kOe (for SCMO).

For the prepared polycrystalline SCSMO, surprisingly, we obtain an unprecedented
magnetoresistance of $\sim10^{15}\%$ at 10 K on application of 90 kOe magnetic field.
This MR value in bulk SCSMO sample is undoubtedly a record value compared with
the previously observed MR in any charge ordered compounds. For example, the MR value
for $La_{0.5}Ca_{0.5}MnO_3$ compound is $10^8\%$ at 57 K for 80 kOe applied field\cite{Gong}.
By substrate induced strain a huge MR of $10^{11}\%$ has also been observed in
$Nd_{0.5}Ca_{0.5}MnO_3$ thin film at 50 K on application of 70 kOe magnetic field\cite{Prellier}.
The charge ordered $Pr_{0.5}Ca_{0.5}MnO_3$ compound shows MR of $10^8\%$ at 60 K on
application of 150 kOe magnetic field due to strain induced by the substrate\cite{Buzin}. Moreover,
thin film made out of our parent compound SCMO shows MR of $10^6\%$ on application of
200 kOe external field at 57 K\cite{Rauwel}. At the same time, MR value of our SCSMO
sample reaches to $10^6\%$ at 50 K only in 30 kOe magnetic field.

We also find an ultra-sharp metamagnetic transition\cite{Ghivelder,Tang,Mahendiran,Rana}
below 10 K. We explain the metamagnetic as well as ultra-sharp metamagnetic transition
using martensitic scenario.
Our Monte Carlo simulations using a two-band double-exchange model show that
A-site disorder suppresses the CE-type phase, but the system remains insulating without
any external magnetic field. In an external magnetic field the inhomogeneous disorder
seeds the ferromagnetic-metallic clusters into the system and as a result resistivity
decreases at low temperatures which gives rise to large magnetoresistance.

%-----------------------------------------------------------------------------------
\section*{Materials and Methods}
%-----------------------------------------------------------------------------------

\subsection*{Sample preparation.}  The polycrystalline
$Sm_{0.5}Ca_{0.25}Sr_{0.25}MnO_3$ compound was prepared by the well known
sol-gel chemistry route. For the sample preparation, high pure (99.99 \%)
Sm$_2$O$_{3}$, CaCO$_3$, Sr(NO$_3$)$_2$, MnO$_2$, oxalic acid and citric acid were used
as the the constituents.
The appropriate amounts of Sm$_2$O$_{3}$ and CaCO$_3$ were converted to their nitrates
forms using concentrated nitric acid (HNO$_3$) and dissolved into millipore water. Since
MnO$_2$ does not dissolve directly in $HNO_3$, hence the appropriate
amount of oxalic acid was used to convert it in oxalate form which dissolve easily into
millipore water in the presence of nitric acid. After preparing the individual clear
solutions of the constituent elements (Sm$_2$O$_{3}$, CaCO$_3$, Sr(NO$_3$)$_2$ and MnO$_2$),
all solutions were mixed up homogenously by using a magnetic stirrer for 30 minutes. For
making the precursor solution, suitable amount of citric acid was added to this homogenous
solution. Extra water was slowly evaporated by continuously heating this solution using a
water bath and maintaining the temperature of the water bath $\sim$80 $^0$C. This slow
evaporation was continued up to the gel formation. After formation of the gel, it was
decomposed at slightly higher temperature ($\sim$150 $^0$C) and the black porous powder
was formed. To evaporate the extra citric acid, porous powder was directly heated at
500 $^0$C for 4 hrs. To get crystalline single phase bulk compound, first the powder
sample was pelletized and heated at 1300 $^0$C for the time period
of ($\sim$36 hrs.)

\subsection*{TEM measurements.}

High-resolution transmission electron microscopy (HRTEM) has been performed using a FEI
Tecnai, F30 microscope with a point to point resolution of 1.8 \AA.  The composition of
the prepared sample was determined from energy dispersive X-ray spectroscopy (EDS)
analysis using an analyzer mounted on the same microscope.

\subsection*{Electrical transport measurement.} Resistance measurements in the absence
of external magnetic field was carried out using Keithley source and measure unit 2651A
by four probe method in the temperature range 20 K to 300 K. The low
temperature measurement of zero field resistivity was extended below 20 K by using Keithley
Electrometer 6517A in the capacitor arrangement method and below 10 K the resistance
value exceeds our measurement limit-I ($10^{13}$ ohm). In-field resistivity measurements
were carried out using 2651A with the measurement limit-II $10^{11}$ ohm.

\subsection*{Magnetic property measurements.} Magnetic properties has been measured
using Superconducting Quantum Interference Device Magnetometer (SQUID-VSM) of Quantum
Design in the temperature range 2 K - 300 K with maximum magnetic field value of 70 kOe.

\subsection*{XRD measurements.} Room temperature powder x-ray diffraction and the
temperature dependent XRD measurements in the temperature range 15 K- 300 K has been
performed using RIGAKU-TTRAX-III diffractometer with rotating anode Cu source of wavelength
$\lambda=1.54A^0$ ($Cu{K_\alpha}$). Measurements has been performed at 9 kW power.

To see the effect of magnetic field on XRD line width broadening, zero field, 70 kOe
infield and followed by zero field measurements were performed using Low temperature and
high magnetic field XRD set up at UGC-DAE CSR (Indore).

\subsection*{Heat capacity measurements.} Magnetic field dependence as well as temperature
dependence of heat capacity of the sample has been carried out using Physical Property
Measurements System of Quantum Design.

%------------------------------------------------------------------------------------------
\section*{Results and Discussion}
\subsection*{Synthesis and structural characterization.}
%------------------------------------------------------------------------------------------

High quality polycrystalline SCSMO was prepared by the well-known sol-gel technique
(see methods for details). The crystalline structure and single phase nature of the
sample was studied using room temperature x-ray diffraction (XRD) and transmission
electron microscopy (TEM). XRD pattern (see Fig.~\ref{TEM}(a)) [analysis is discussed
in supplementary section I(A)] and TEM diffraction pattern (see Fig.~\ref{TEM}(e))
shows the orthorhombic crystallographic symmetry (Pnma) of the sample. Room
temperature chemical analysis using energy dispersive spectroscopy (EDS) measurement
[see supplementary section I(C)] shows that the elements are distributed
homogeneously with stoichiometricity.

We also present XRD analysis at several low temperatures in Fig.~\ref{TEM}(b)-(d).
With lowering of temperature another crystallographic phase with monoclinic symmetry
P21/m along with the room temperature symmetry Pnma i.e. combinations of Pnma and
P21/m appears below ${T \leq 120K}$. The evolution of the appeared new peak
corresponding to the P21/m monoclinic space group [see Fig.~\ref{TEM}(c)]. It
indicates that below 120 K the new peak appears at $2\theta=47.5^0$ and its intensity
remains almost constant below 100 K. It implies that below 100 K, P21/m phase gets
trapped within host Pnma phase and creates lattice strain in the host Pnma phase.
Strain is also evident from the TEM analysis [see Fig.~\ref{TEM}(f)] at 100 K.
It is earlier shown that the antiferromagnetic transition is associated with
structural transition from orthorhombic to orthorhombic$+$monoclinic phase at 135 K
in $Sm_{0.5}Sr_{0.5}MnO_3$\cite{KurbakovPRB,KurbakovJPCS}. So, the monoclinic structure
favors the antiferromagnetic phase in $Sm_{0.5}Sr_{0.5}MnO_3$. In our case, the SCSMO
compound also undergoes the same kind of structural transition at 120 K. We will show
in the next section that the system also undergoes an
antiferromagnetic transition at 120 K. Thus, the monoclinic P21/m phase
[shown in Fig.\ref{TEM} (b) and (c)] is associated with the CE-AFM phase.

%***************************************************************************************
\begin{figure*}
\includegraphics[width=0.95\textwidth]{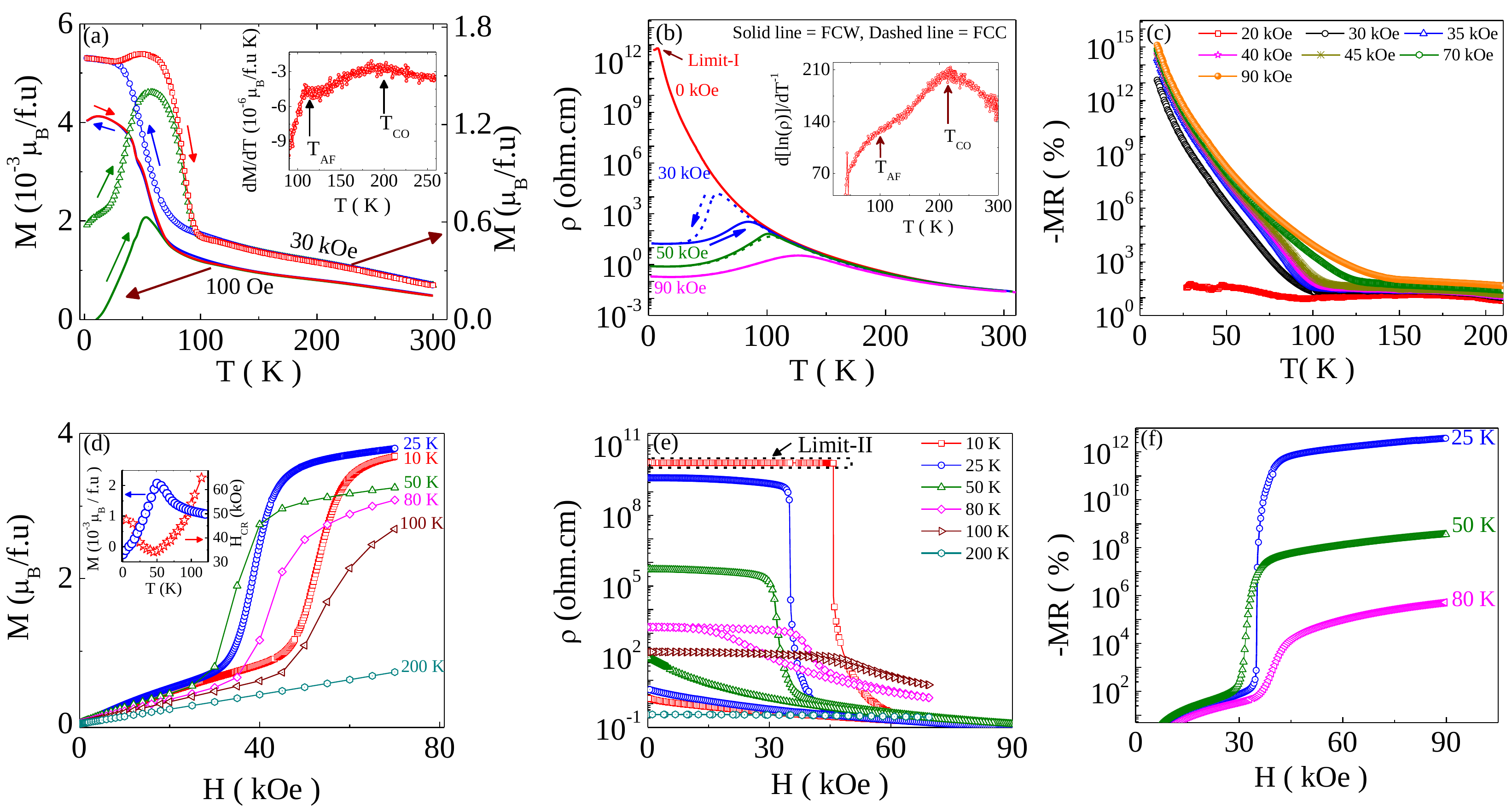}
\centering \caption[ ]
{\label{MR_T} Magnetotransport properties:
(a) Variation of magnetization with temperature in 100 Oe and 30 kOe external magnetic
field. The inset shows the signature of CO and AFM ordering, indicated by arrows, from
the temperature derivative of M(T) data taken in FCW protocol in the presence of 100 Oe
magnetic field.
(b) Temperature dependent resistivity without (red), with [30 kOe (blue), 50 kOe (olive),
90 kOe (violet)] external magnetic fields. The dotted lines represent the resistivity data
taken during field cooling cycle and the solid lines are for the FCW cycle. The inset
shows variation of activation energy (${E_A\propto d[ln(\rho)]/dT^{-1}}$) with temperature,
calculated from temperature dependence of zero field resistivity data and ordering
temperatures are indicated by arrows.
(c) Magnetic field dependence of MR with temperature for different magnetic fields.
(d) Magnetization vs. magnetic field and (e) resistivity vs. magnetic field
at different temperatures. Inset in (d) shows the temperature dependence of $H_{CR}$ and ZFC
magnetization in 100 Oe magnetic field. (f) Magnetic field dependent magnetoresistance
at three different temperatures. See materials and methods section for
explanation of limit-I and limit-II, mentioned in (b) and (e) respectively.}
\end{figure*}
%***************************************************************************************

The effect of external magnetic field (70 kOe) on the XRD line width broadening at 2 K
[see Fig.~\ref{TEM} (d)] indicates the decrease of the line width (full width at
half maxima) at $2\theta = 47.5^0$ from $0.58^0$ to $0.46^0$ and remains at the same
value even after removing the field. This shows that strain decreases in an external
magnetic field and stays as it is even after removing the field.

%------------------------------------------------------------------------------------------
\subsection*{Magnetic and Magnetotransport Measurements}
%------------------------------------------------------------------------------------------
The temperature dependent magnetization and resistivity for different applied
magnetic fields are shown in Figs. \ref{MR_T}(a) and (b). Interestingly below $60K$
magnetization increases, but ferromagnetic fraction is very small for
100 Oe magnetic field and this fact is reconfirmed from the thermoremanent
magnetization measurements [discussed in supplementary section I(D)]. As a result,
the system without any magnetic field remains insulating at low temperature
as seen in Fig.~\ref{MR_T}(b), but the stability of the CO-AFM state
decreases (discussed in the next paragraph). This is unlike our previously studied
core-shell (ferromagnetic core and antiferromagnetic shell) nanostructures~\cite{Das2}
for which a comparatively small resistivity ($\sim$ 10$^{2}$ Ohm-cm) was obtained at
low temperatures. Below a certain temperature ${(T < 10 K)}$ the value of
resistance is ${R\sim 10^{13}ohm}$ for SCSMO, which is limiting
value of our measuring instruments (see method section for details). The ordering
temperatures ($T_{CO}$ and $T_N$) are indicated in the inset of
Figs.~\ref{MR_T} (a) and (b) and similar $T_{CO}$ is also obtained from the
heat capacity measurement [please see  supplementary section I(E)].

A 30 kOe magnetic field induces larger ferromagnetic fraction and as a result resistivity
decreases considerably (from $10^{13}$ Ohm-cm to ${\rho \sim 17.5}$ Ohm-cm
at T = 2.5 K) at low temperatures [see Figs. 2(a) and (b)]. This shows
that not only $T_{CO}$ decreases, but the robustness of CO-AFM state is reduced in SCSMO.
A huge hysteresis is observed between field cooling and heating cycle for 30 kOe in the
temperature range ${50 K < T < 100K}$, in both magnetization and resistivity curves, which
is the signature of field induced electronic phase separation\cite{dagottoE}. With further
increase of applied magnetic field this coexistence is suppressed and low temperature
resistivity goes to $10^{-2}$ Ohm-cm (for 90 kOe).

To quantify the field induced change in resistivity, we plot the MR
(${= (\rho(H)-\rho(0))/\rho(H)\times 100 }$) with temperature at different magnetic field
in Fig.~\ref{MR_T}(c). MR is $\sim$10$\%$ for 20 kOe, but astonishingly increases to $10^{13}\%$ for
30 kOe external magnetic field and as large as ${\sim 10^{15}\%}$ at 10 K for
90 kOe magnetic field. The magnetoresistance value as high as $10^{15}\%$ in polycrystalline
stable SCSMO compound is a unique observation.

%***************************************************************************************
\begin{figure*}
\includegraphics[width=0.9\textwidth]{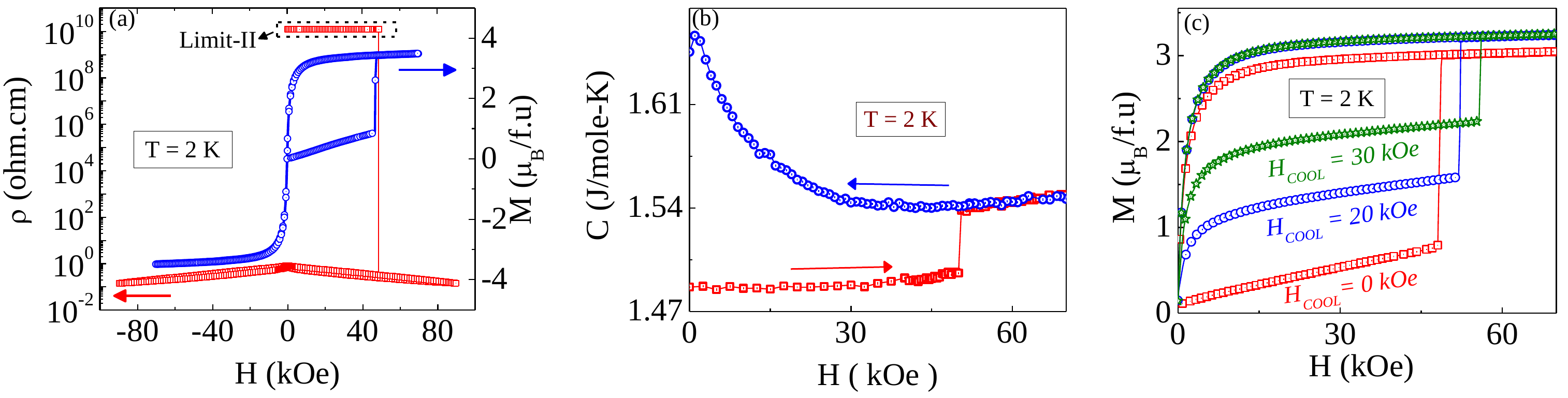}
\centering \caption[ ]
{\label{UMT} Ultra-Sharp metamagnetic transition: (a) Magnetization (and resistivity) vs
magnetic field at 2 K. See materials and methods section for explanation of
limit-II. (b) Variation of heat capacity with external magnetic field at 2 K. Here red
and blue symbols are for the C(H) data taken during increasing and decreasing field
respectively. (c) Effect on isothermmal magnetization with different cooling fields at 2 K.
}
\end{figure*}
%***************************************************************************************

Isothermal magnetization measured in ZFC protocol at different temperatures [presented in
Fig.~\ref{MR_T}(d)] shows the field induced metamagnetic transitions. Experimentally obtained
saturation magnetic moment (${3.82\mu_B}$) at 30 K is close to the estimated magnetic moment
for the full saturation of the $Mn^{3+}$/$Mn^{4+}$ and $Sm^{3+}$ ions
(${3.85\mu_B}$). It clearly indicates that the CO-AFM state melts completely on application
of the magnetic field via metamagnetic transition near critical magnetic field $H_{CR}$.
Temperature variation of $H_{CR}$ measured from M(H) isotherms
is opposite to that of ZFC magnetization as shown in the inset of Fig.~\ref{MR_T}(d).
The FM components present in the ZFC sample act as the nucleation
center and grows at the expense of AFM components in an external magnetic field resulting
a metamagnetic transition at $H_{CR}$. As the ZFC magnetization increases from 2 K to 50 K,
enhanced thermal energy reduces the $H_{CR}$ from 47.8 kOe at 2 K to 34.5 kOe at 50 K.
For $T > 50 K$ the
$H_{CR}$ increases due to reduction of ZFC magnetization. In the absence of AFM components,
above $T_N$, the field induced metamagnetic transition vanishes.

Fig.~\ref{MR_T}(e) shows the resistivity vs. magnetic-field isotherms measured at different
temperatures. Interestingly, the isotherms measured at lower temperature (10 K) show
exceptionally sharp metamagnetic jump. With increasing temperature, the loop area reduces
between increasing and decreasing field sweeps and at 100 K it almost disappears. Thus,
we believe that field induced phase coexistence sustains up to $\sim$120 K which is the
antiferromagnetic ordering temperature. Another point to note here is that the system
remains in the low resistive state even after removing the field for $T < 50 K$.
%Also the ZFC magnetization shows a peak around 50 K
Due to resistivity measurement limitation the resistivity is limited to $10^{10}$ ohm-cm
in an external magnetic field and as a result we only able to measure the MR for temperature
above 25 K. The MR is equal to $10^{12}\%$ ($10^{4}\%$) at 25 K (80 K) for 45 kOe magnetic
field as shown in Fig.~\ref{MR_T}(f) and will be even larger for lower temperatures.

%----------------------------------------------------------------------------------
\subsection*{Ultra-Sharp Metamagnetic Transition}
%----------------------------------------------------------------------------------

Interestingly, the resistivity (and magnetization) vs. magnetic-field isotherms,
[see Fig.~\ref{UMT}(a)], at 2 K  show exceptionally ultra-sharp steps with a width
of $\sim$10 Oe (the smallest step kept during the measurement is 3 Oe). To explore
the origin of this ultra-sharp metamagnetic transition we turn now to measure
the heat capacity (C) with magnetic field. During the field increasing cycle the
heat capacity data show a step-like behavior around 48 kOe, as shown in Fig.~\ref{UMT}(b),
and has one-to-one correspondence with the step-like behavior observed in the magnetization
and resistivity isotherms. This is opposite to the case observed in earlier studies
where C decreases sharply with H at the transition point and was associated with the
rise in temperature of the system due to release of energy that assists the abrupt
field-induced transition\cite{Ghivelder,Maji}. The sharp increase in the heat capacity
at 48 kOe magnetic field rules out the possibility of exothermic temperature driven
avalanche metamagnetic transition observed in our system and we believe that the system
changes martenisitically.% as discussed in Fig.~\ref{MR_H}(c).

To ascertain the martensitic nature we also study the isothermal magnetization (at 2 K)
for different cooling fields [see Fig.~\ref{UMT}(c)]. With increasing
cooling field FM fraction increases and as a result interfacial elastic energy increases.
Due to this the critical field increases with increasing the cooling field. At the
same time, the critical field decreases slightly when we increase the sweep rate from
10 Oe/sec to 200 Oe/sec for ZFC sample [see supplementary section I(F)
for details]. This is because the lattice has adequate time to accommodate the
induced interfacial strain between AF and FM domains for lower sweep rate and a larger
magnetic field is required to break the energy barrier. Both these scenario indicates
the martensitic nature of the transition\cite{Maji,Lion,Wu,Shankaraiah,Hardy}.
Our resistivity relaxation study [see supplementary section I(G)] also
confirms the martensitic nature of the metamagnetic transitions
[Fig.~\ref{MR_T}(d) and (e)] at higher temperatures.

%%%%%%%%%%%%%%%%%%%%%%%%%%%%%%%%%%%%%%%%%%%%%%%%%%%%%%%%%%%%%%%%%%%%%%%%%%%%%%%%%%%%%%
\subsection*{Theoretical Simulation}
%%%%%%%%%%%%%%%%%%%%%%%%%%%%%%%%%%%%%%%%%%%%%%%%%%%%%%%%%%%%%%%%%%%%%%%%%%%%%%%%%%%%%%

In this section we discuss the physical origin of the huge magnetoresistance
using a two-band double exchange model including super-exchange ($J$) and electron-phonon
coupling ($\lambda$). Our model Hamiltonian~\cite{dagotto,yunoki,pradhan1,anamitra,pradhan}
(see Supplementary Section II for details), effectively a lattice of Mn ions,
qualitatively reproduces the phase diagram of manganites. For SCMO (and SSMO) like
materials, involving two A-type elements~\cite{Tokura2}, one generally add
$\sum_i \epsilon_i n_i$ such that $\overline{\epsilon}_j=0$
to model the A-site cationic disorder~\cite{dagotto,pradhan}. Considering the fact
that the ${Sr^{2+}}$ ions occupy randomly in the A-site in the polycrystalline SCSMO
compound and being larger in size compared with both $Sm^{3+}$ and $Ca^{2+}$
creates chemical disorder. This is also evident from the HRTEM images at room
temperature, shown in Supplementary Section I(B) [Figs. 6(b) and (c)]. So in order to model
SCSMO, we neglect the disorder between Sm and Ca elements and incorporate the Sr
disorder by adding $\sum_i \epsilon_i n_i$ at each Mn site picked from the distribution
$P(\epsilon_i) = {1 \over 4} \delta(\epsilon_i - \Delta) + {3 \over 4} \delta(\epsilon_i + \Delta)$,
where $\Delta$ is the quenched disorder potential. We add a Zeeman coupling term
$-\sum_i {\bf h}\cdot{\bf S}_i$ to the Hamiltonian in an external magnetic field,
where ${\bf S_{\rm i}}$ are Mn t$_{\rm 2g}$ spins to analyze the magnetoresistance. We
measure $J$, $\lambda$, $\Delta$, $h$ and temperature ($T$) in units of kinetic hopping
parameter $t$. The estimated value of $t$ in manganites is 0.2 eV~\cite{dagotto}.

%%%%%%%%%%%%%%%%%%%%%%%%%%%%%%%%%%%%%%%%%%%%%%%%%%%%%%%%%%%%%%%%%%%%%%%%%%%%%%%%%%%%%
\begin{figure} [t]
\centerline{
\includegraphics[width=9.5cm,height=8.5cm,clip=true]{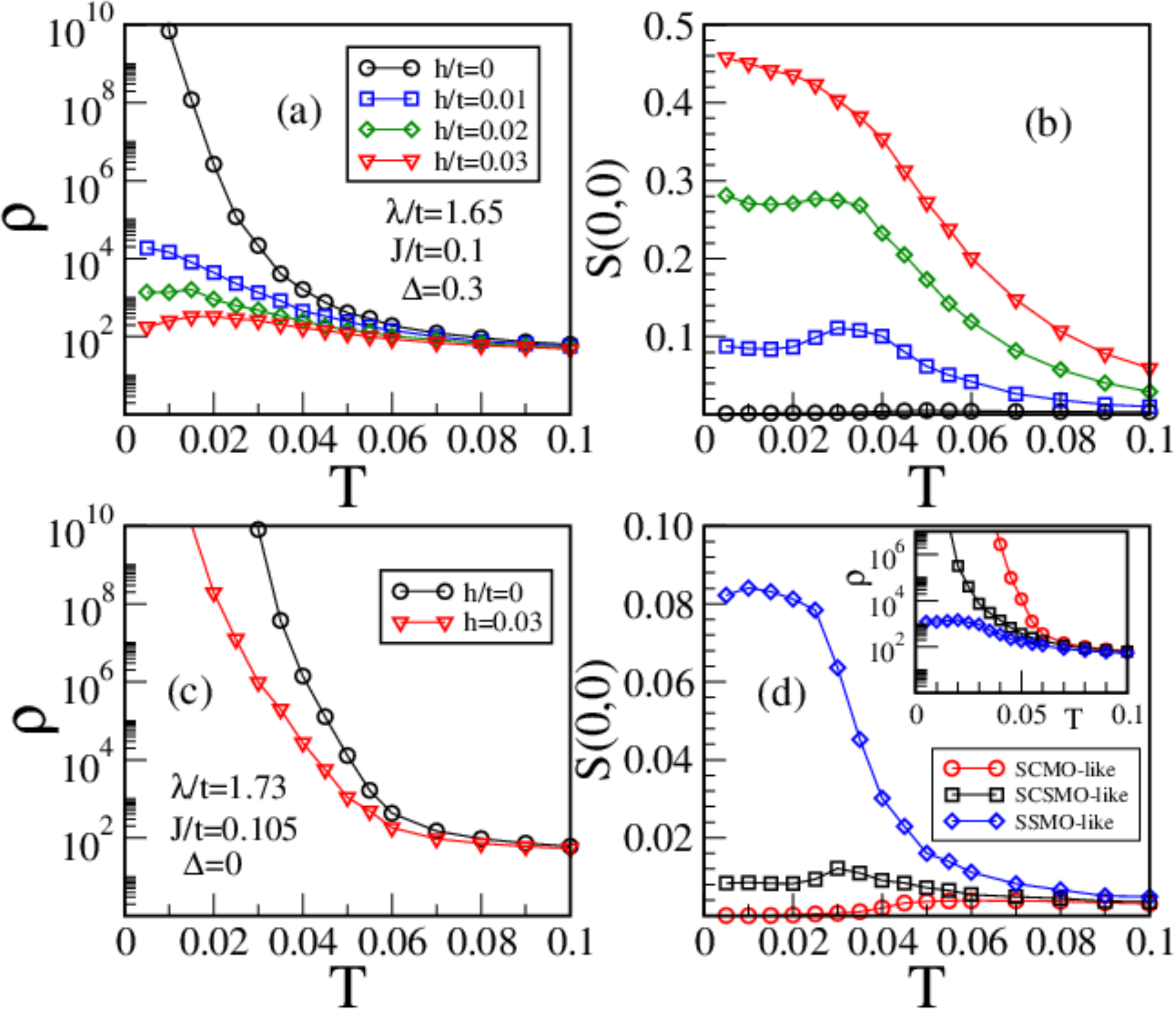}}
\caption{Temperature dependence of (a) the resistivity $\rho$ in units of
${\hbar a}/{\pi  e^2 }$ and (b) the FM structure factor S(0,0) in different external
magnetic field $h/t$ values
for $\lambda/t$= 1.65, $J/t$ = 0.1 and $\Delta/t$ = 0.3 (SCSMO-like materials).
Legends in (a) and (b) are the same. Electron density is fixed at $n$ = 0.5 in all
figures. (c) Temperature dependence of $\rho$ for $\lambda/t$= 1.73, $J/t$ = 0.105 and
$\Delta/t$ = 0 (SCMO-like materials). (d) Temperature dependence of FM structure factor
S(0,0)[inset: resistivity] for three sets of parameters (mimicking SCMO, SCSMO and SSMO)
in a very small magnetic field $h$=0.002. See the text for details.}
\label{fig:th_1}
\end{figure}
%%%%%%%%%%%%%%%%%%%%%%%%%%%%%%%%%%%%%%%%%%%%%%%%%%%%%%%%%%%%%%%%%%%%%%%%%%%%%%%%%%%%%

A spin-fermion Monte Carlo (MC) technique based on the travelling cluster
approximation~\cite{tca-ref} (TCA) is used on a two dimensional 24$\times$24 lattice
(see supplementary section II for details). We use $J/t$ = 0.1 and $\lambda/t$ = 1.65
that reproduces the CE-CO-OO-I phase~\cite{pradhan1} at electron density $n$ = $1-x$ = 0.5.
The electron density is the number of itinerant $e_{\rm g}$ electrons per Mn site in
our calculations. The system, as shown in Figs.~\ref{fig:th_1}(a) and (b), remains
insulating at low temperatures for $\Delta/t$ = 0.3 and $h$ = 0 and the ferromagnetic
structure factor S(0,0) is $\sim$0.001 (for an outline of the resistivity and
the magnetic structure factor calculations please see Supplementary Section II).

The resistivity at low temperatures decreases with magnetic field $h$
[Fig.~\ref{fig:th_1}(a)] similar to our experimental results. This is due to the
increase of the FM correlations at low temperatures [Fig.~\ref{fig:th_1}(b)].
On the other hand, SCMO-like materials [Fig.~\ref{fig:th_1}(c)] remains insulating at
all temperatures even for $h$ = 0.03. Recall that SCMO (SSMO) has smaller (larger)
bandwidth than SCSMO. In our model calculations larger $\lambda/t$ (and $J/t$)
corresponds to smaller bandwidth or vice versa. For clarity we use $\Delta/t$ = 0
(due to the small mismatch between Sm and Ca ionic radii) and set $\lambda/t$= 1.73,
$J/t$ = 0.105 for SCMO-like materials. For SSMO-like materials we set
binary disorder~\cite{dagotto,pradhan} with $\Delta/t$= 0.3
and use $\lambda/t$= 1.57, $J/t$ = 0.095. The FM correlations
at low temperatures increases and the resistivity decreases from SCMO-like to
SCSMO-like to SSMO-like materials [Fig.~\ref{fig:th_1}(d)] similar to
the experimental results qualitatively~\cite{Tokura2,Tomioka}.

%%%%%%%%%%%%%%%%%%%%%%%%%%%%%%%%%%%%%%%%%%%%%%%%%%%%%%%%%%%%%%%%%%%%%%%%%%%%%%%%%%%%%
\begin{figure} [t]
\centerline{
\includegraphics[width=9.0cm,height=8.5cm,clip=true]{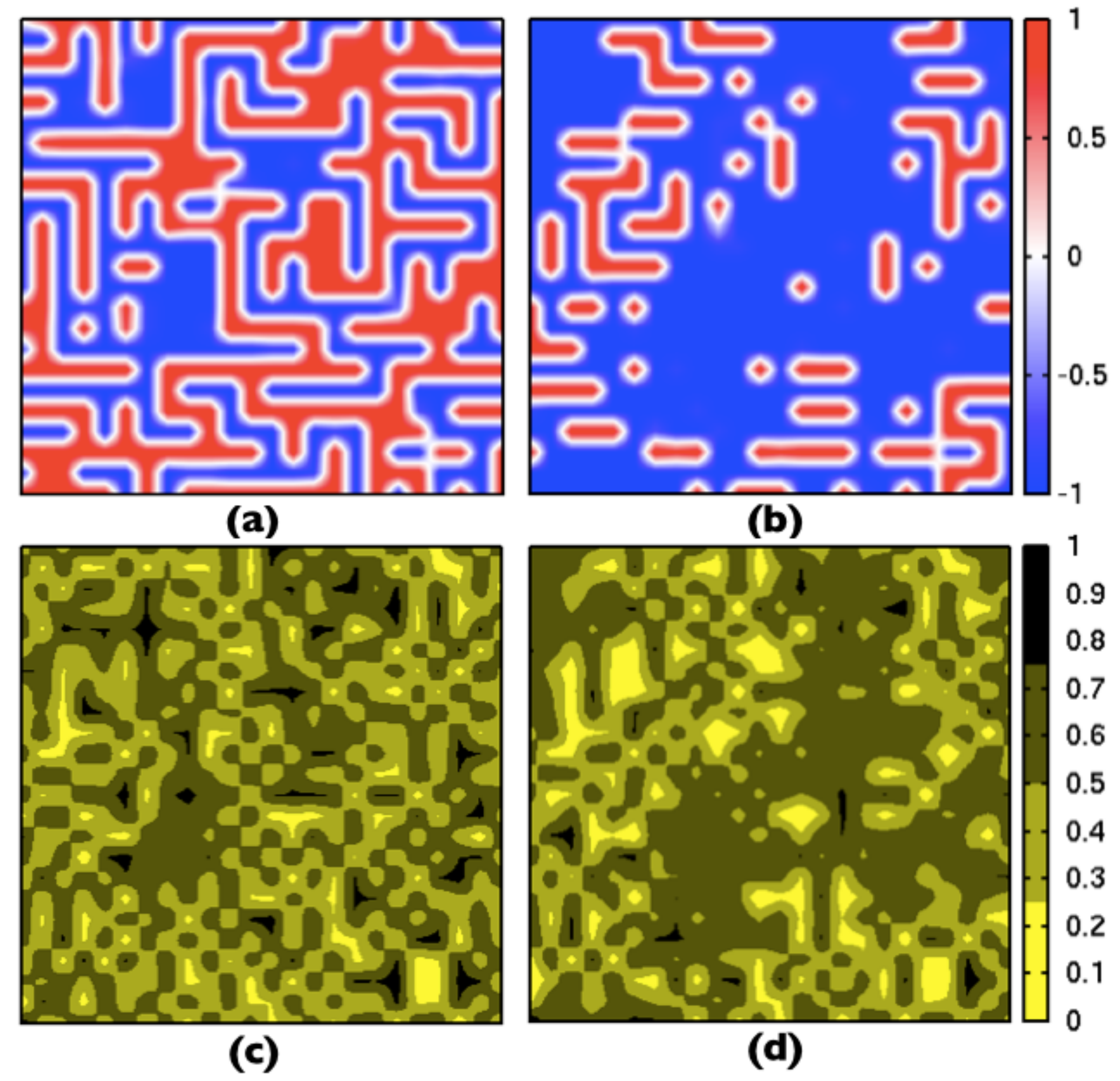}}
\caption{(a) and (b) The $z$ components of simulated Mn (t$_{\rm 2g}$) spins;
(c) and (d) electron density for each site on a 24$\times$24 lattice
at $T$ = 0.01 using $\lambda/t$= 1.65, $J/t$ = 0.1 and $\Delta/t$ = 0.3.
In (a) and (c) $h/t$ = 0.002, and in (b) and (d) $h/t$ = 0.02.
}
\label{fig:th_2}
\end{figure}
%%%%%%%%%%%%%%%%%%%%%%%%%%%%%%%%%%%%%%%%%%%%%%%%%%%%%%%%%%%%%%%%%%%%%%%%%%%%%%%%%%%%%

We turn now to
understand the magnetoresistance in SCSMO by plotting Monte Carlo snapshots obtained at
$T$ = 0.01. For $h$ = 0.002 the system remains insulating
[see Figs.~\ref{fig:th_1} (d)] due to the CE-type correlations
without any significant ferromagnetic charge disordered regions [see Figs.~\ref{fig:th_2}
(a) and (c)] in the system. For $h$ = 0.02, ferromagnetic clusters coexist with zig-zag
ferromagnetic chains and the electron density is roughly homogeneous ($\sim$0.65)
within the ferromagnetic clusters [see Figs.~\ref{fig:th_2} (b) and (d)]. Moreover, the
ferromagnetic clusters in SCSMO get connected with each other at reasonable large magnetic
fields unlike SCMO-like materials for which resistivity decreases at
lower temperatures. So, overall the disorder due to Sr ions present in SCSMO weakens
the CE-type state of SCMO and seeds the ferromagnetic charge-disordered clusters,
but remains insulating at low temperatures. In an external magnetic field ferromagnetic
clusters grow and get connected that give rise to a large magnetoresistance in SCSMO
samples.

%----------------------------------------------------------------------------------
\section*{Conclusion}
%----------------------------------------------------------------------------------

In summary, by controlling electronic phase separation in narrow band width CO-AFM
material a record value of magnetoresistance ($\sim10^{13}\%$ in a 30 kOe external
magnetic field and $\sim10^{15}\%$ in 90 kOe external magnetic field at 10 K ) till date is
obtained in SCSMO polycrystalline compound. The observed huge value of
magnetoresistance has been explained by model Hamiltonian calculations. We also observed
magnetic field induced ultra-sharp meta-magnetic transition at low temperatures.
The many order of magnitudes higher MR than any other magnetoresistive materials
reported so far will motivate experimenters to design efficient magnetic
materials for future applications.

%----------------------------------------------------------------------------------
\section*{Conflict of Interest}
%----------------------------------------------------------------------------------
The authors declare no conflict of interest.

%------------------------------------------------------------------------------------
\section*{Acknowledgements}
Sanjib Banik would like acknowledge Department of Atomic Energy (DAE), Govt. of India
for finalcial support. K. P. acknowledges the use of TCMP computer cluster at SINP.
The authors would like to sincere thanks to P. B. Pal and
M. G. Mustafa of SINP for useful discussions. The work at SINP has been supported through
CMPID-DAE project.

\section*{Author information}
I.D. developed the concept of the study. SB and KD prepared the samples. S.B., K.D.,
and I.D. performed the experimental works and conducted the data analysis. N.P.L did the
field dependent XRD measurements, B.S carried out the TEM measurements, and K.P. performed
the numerical calculations. S.B., K.D., T.P., K.P., and I.D. wrote the draft of the paper
and all authors reviewed the manuscript.

%------------------------------------------------------------------------------------

%--------------------------------------------------------------------------------
%\section*{References}
%--------------------------------------------------------------------------------

\newpage
\section*{supplementary information}

\section*{I. Experimental Work}
\subsection{X-ray diffraction study}

The room temperature XRD data has been profile fitted using Pnma space group
and shows the single phase nature of the sample with orthorhombic structure.
The low temperature data has also been tried to fit with single Pnma space group
but does not give satisfactory fitting. Previously it was
reported that $Sm_{0.5}Sr_{0.5}MnO_3$ sample undergoes crystallographic phase
coexistence at low temperature with monoclinic P21/m symmetry\cite{Kurbakov}.
Since the present sample is in between $Sm_{0.5}Ca_{0.5}MnO_3$ and
$Sm_{0.5}Sr_{0.5}MnO_3$, we had tried to fit the low temperature (15 K) XRD data
with P21/m but it also does not give any satisfactory fitting. Finally, fitting
with Pnma+P21/m gives the best fit. We present the extracted
lattice parameters in Table. I.

%**********************************************************************************
\begin{figure} [h]
\includegraphics[width=0.45\textwidth]{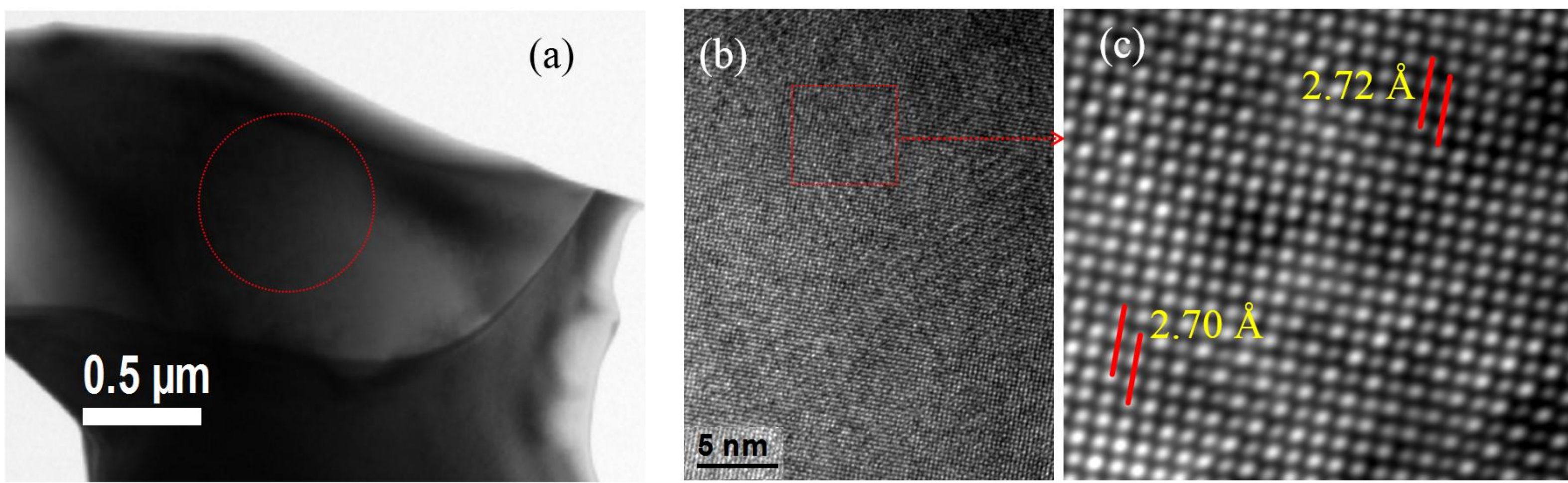}
\centering
\caption[]{
Panel (a) gives a typical TEM overview of one crystallite where two grains are visible.
Panel (b) gives a typical HRTEM overview of one crystallite. Panel (c) [a magnified
view of tiny domains in (b)] displays an example of atomic scale variation of the contrast.}
\end{figure}
%**********************************************************************************

\subsection{HRTEM study}

A typical TEM image of a crystallite of the sample has been presented in Fig. 6(a). In the marked
portion electron diffraction pattern has been taken which has been shown in the Fig. 1 (e) of the
main text. The HRTEM image of the sample at room temperature is presented in Fig. 6(b) and Fig. 6(c)
is the magnified view of the small domain. It shows the change in contrast due to the segregation
of different atoms in the atomic scale. The interplanar spacing has been shown in the figure. Here
another point needs to mention that the error bar in determining the interplanar spacing in the
figure is $\pm 0.04 \AA$~\cite{Satpati}.

%**********************************************************************************
\begin{figure}[h]
\includegraphics[width=0.45\textwidth]{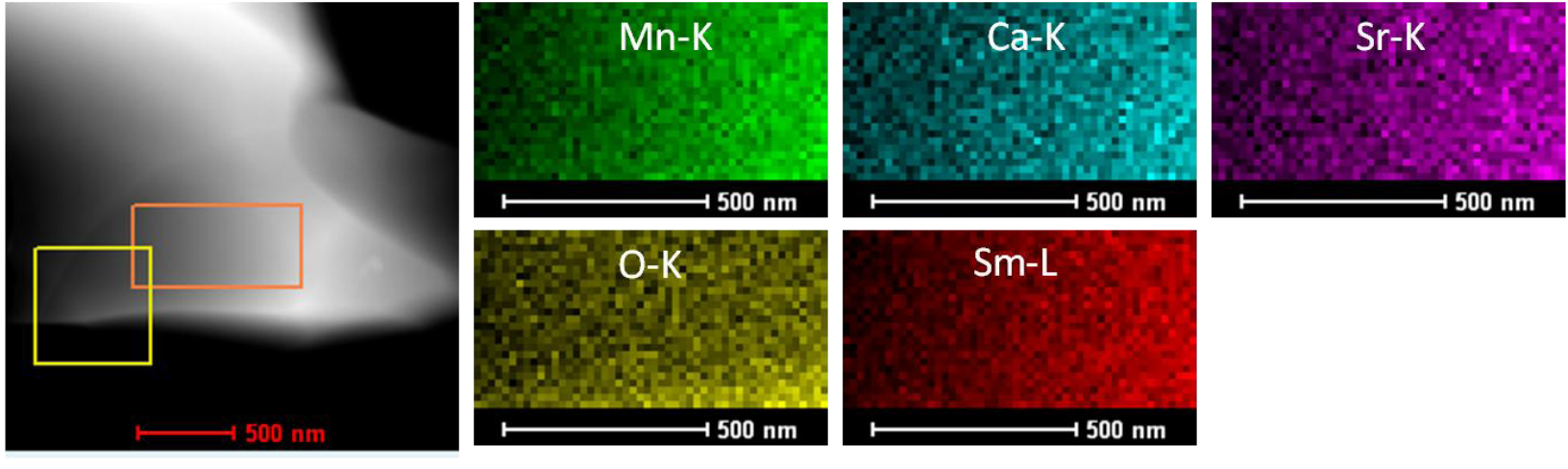}
\centering
\caption[]{STEM-HAADF image and corresponding drift corrected chemical maps from the area marked by orange box in left panel.}
\end{figure}
%**********************************************************************************

%**********************************************************************************
\subsection{EDS Analysis}
%**********************************************************************************

The EDS analysis in scanning transmission electron microscopy high angle annular
dark field (STEM-HAADF) mode was carried out with a $\sim$2 nm probe and shows
that the crystal is having highly homogeneous distribution of elements [Fig. 7].
The chemical analysis (see Table. II) confirms the stoichiometric nature
of the compound.

\begin{table*}
\caption{The lattice parameters and unit cell volumes of the sample ${Sm_{0.5}Ca_{0.25}Sr_{0.25}MnO_3}$}
\centering
\begin{tabular}{@{\hskip 0.18in}c @{\hskip 0.18in}c @{\hskip 0.18in}c @{\hskip 0.18in}c @{\hskip 0.18in}c @{\hskip 0.18in}c @{\hskip 0.18in}c} % used for centering columns
\hline\hline                                                                                                                     %inserts double horizontal line
Temperature (K)  &  Space group    &    a ($\AA$)    &   b ($\AA$)   &   c ($\AA$)   &  V ($\AA^3$)   &                  \\ [0.5ex]   %inserts table heading
\hline                                                                                                                           % inserts single horizontal line
300            &   Pnma            &   5.395   &   7.611    &  5.408   &     222.049     &                         \\            % insert body of the table
15             &   Pnma + P21/m    &   5.456   &   7.618    &  5.384   &     223.804     &     Pnma                \\           % insert body of the table
15             &   Pnma + P21/m    &   5.380   &   10.849   &  7.569   &     441.802     &     P21/m               \\ [1ex]     % [1ex] adds vertical space                                \hline
\end{tabular}
\label{LATTICE PARAMETERS}
\end{table*}

\begin{table}[h]
\caption{The EDS analysis of the sample ${Sm_{0.5}Ca_{0.25}Sr_{0.25}MnO_3}$}
\centering
\begin{tabular}{@{\hskip 0.18in}c   @{\hskip 0.18in}c} % used for centering columns
\hline\hline                                                        %inserts double horizontal line
Element  &  Atomic \%                                   \\ [0.5ex]   %inserts table heading
\hline                                                              % inserts single horizontal line
O (K)            &   59.39                              \\           % insert body of the table
Ca (K)           &   4.94                               \\           % insert body of the table
Mn (K)           &   20.43                              \\           % insert body of the table
Sr (K)           &   5.71                               \\           % insert body of the table
Sm (L)           &   9.50                               \\ [1ex]     % [1ex] adds vertical space
\hline
\end{tabular}
\label{EDS analysis}
\end{table}

\subsection{Thermoremanent magnetization measurements}

%**********************************************************************************
\begin{figure}[h!]
\includegraphics[width=0.45\textwidth]{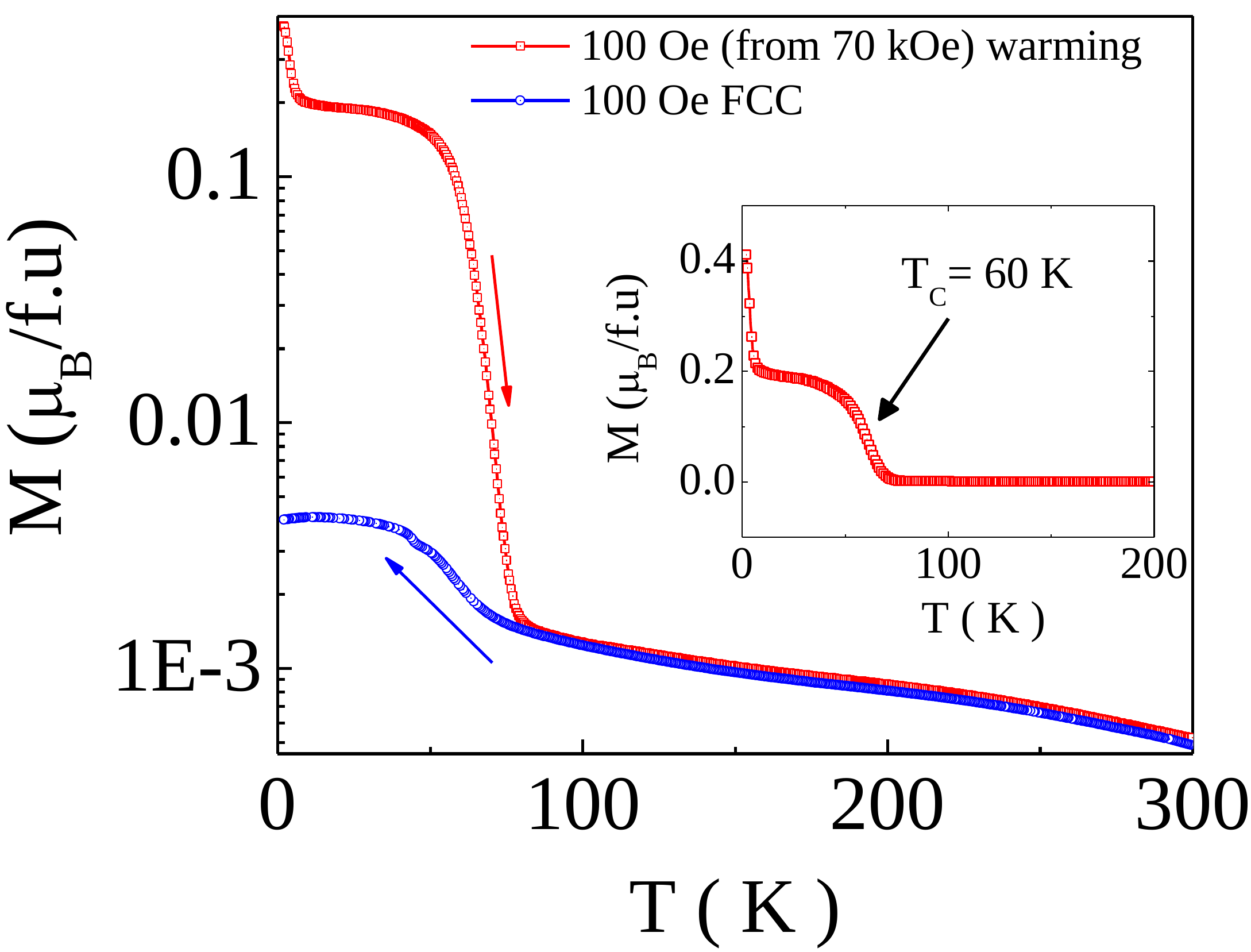}
\centering
\caption[]{Magnetization as a function of temperature measures at H = 100 Oe
external magnetic field. Blue curve is the magnetization during cooling of the
sample (at H = 100 Oe). Red curve (linear scale in the inset) is magnetization during
warming in the presence of H = 100 Oe external magnetic field (before starting
measurements (at T = 2 K) the field cycled from 100 Oe to 70 kOe to 100 Oe).}
\end{figure}
%**********************************************************************************

Thermoremanent magnetization measurements were carried out by the following protocol:
the sample was cooled down from the room temperature in the presence of 100 Oe external
magnetic field and the magnetization data was recorded (denoted by 100 Oe FCC in Fig. 8).
The 100 Oe FCC data indicates the presence of very small fraction of ferromagnetic phase.
After reaching the specified temperature (T = 2 K), magnetic field increases from 100 Oe
to 70 kOe and stayed for some time (approximately 5 mins). After that magnetic field was
reduced to 100 Oe and magnetization data was recorded during warming cycle (see Fig. 8).
During the warming cycle the system changes to a paramagnetic phase at 60 K.
\subsection{Heat Capacity Measurements}

%**********************************************************************************
\begin{figure}[h!]
\includegraphics[width=0.45\textwidth]{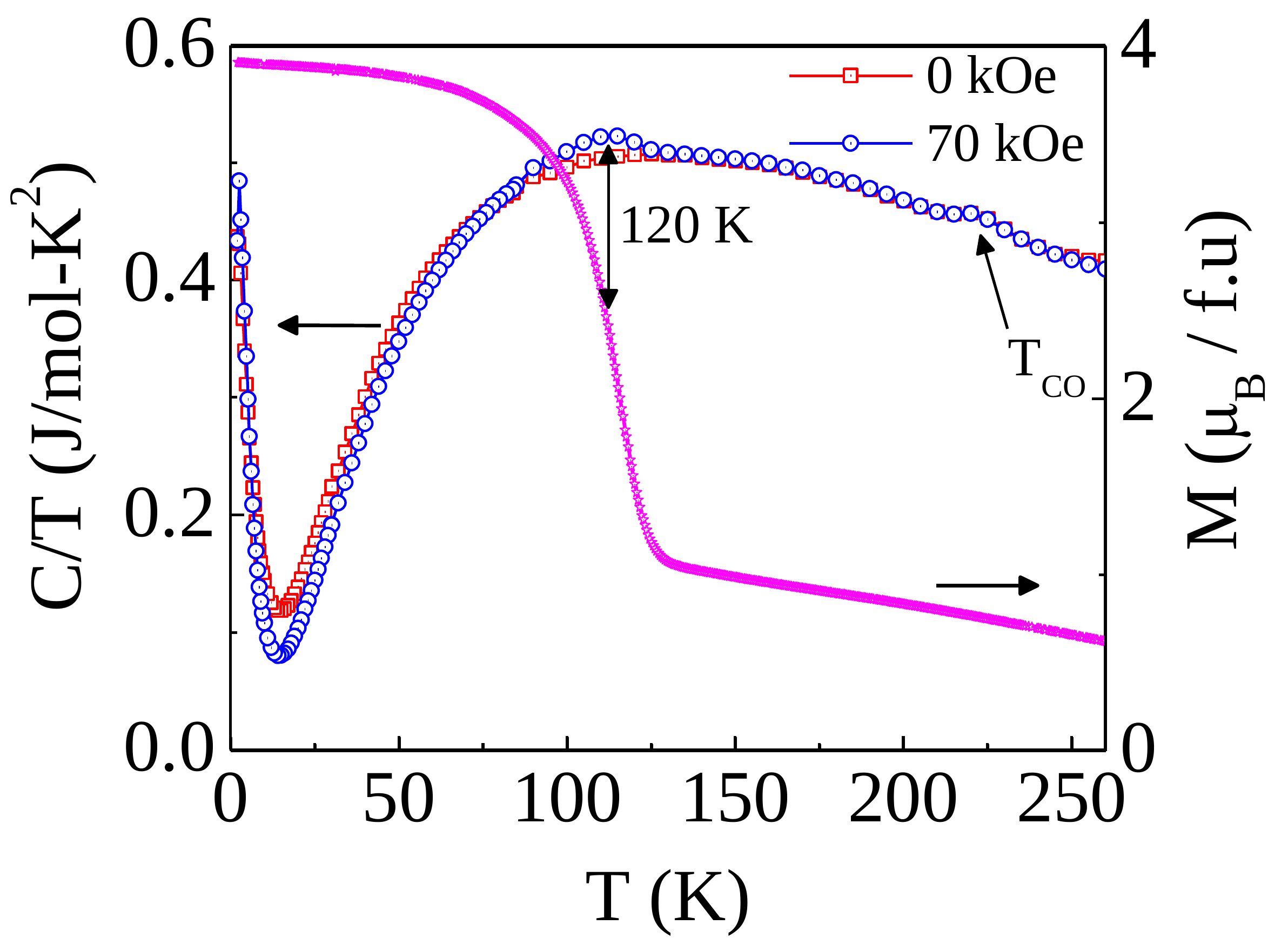}
\centering
\caption[]{Left axes: C/T as a function of temperature (C is heat capacity)
in the absence and in the presence of 70 kOe external magnetic field. Right
axes: Temperature dependence of the magnetization measured in presence of
70 kOe magnetic field.}
\end{figure}
%**********************************************************************************

Around 120 K, magnetization data measured at 70 kOe field also shows a sharp increase
like paramagnetic to ferromagnetic transition [see Fig.9]. Thus the
kink at 120 K in the heat capacity is associated with the paramagnetic to ferromagnetic
transition. Though in the virgin sample around 120 K there is an antiferromagnetic
transition as observed from the derivative of M(T) data in 100 Oe field. Therefore, it can
be concluded that 70 kOe field converts the AFM fraction to FM phase which results in the
rise in the peak in C/T versus T data (see Fig. 9). Additionally, the sharper increasing nature of
C/T below 15 K is observed which is due to the magnetic ordering of $Sm^{3+}$ ions\cite{Tomy,Midya}.

\subsection{Sweep rate dependent magnetization}

Isothermal magnetization (at 2 K) has been measured in ZFC mode
for different field sweep rate (10 Oe/sec and 200 Oe/sec) and the corresponding plot is
presented in Fig. 10. For lowest sweep rate (10 Oe/sec) jump occurs at higher field
compared to that of the maximum sweep rate. For lower sweep rate lattice has adequate
time to accommodate the induced interfacial strain between AF and FM domains but for
higher sweep rate strain propagates rapidly and conversion to FM phase take place.
This scenario indicates the martensitic nature of the transition.

%**********************************************************************************
\begin{figure}[h!]
\includegraphics[width=0.45\textwidth]{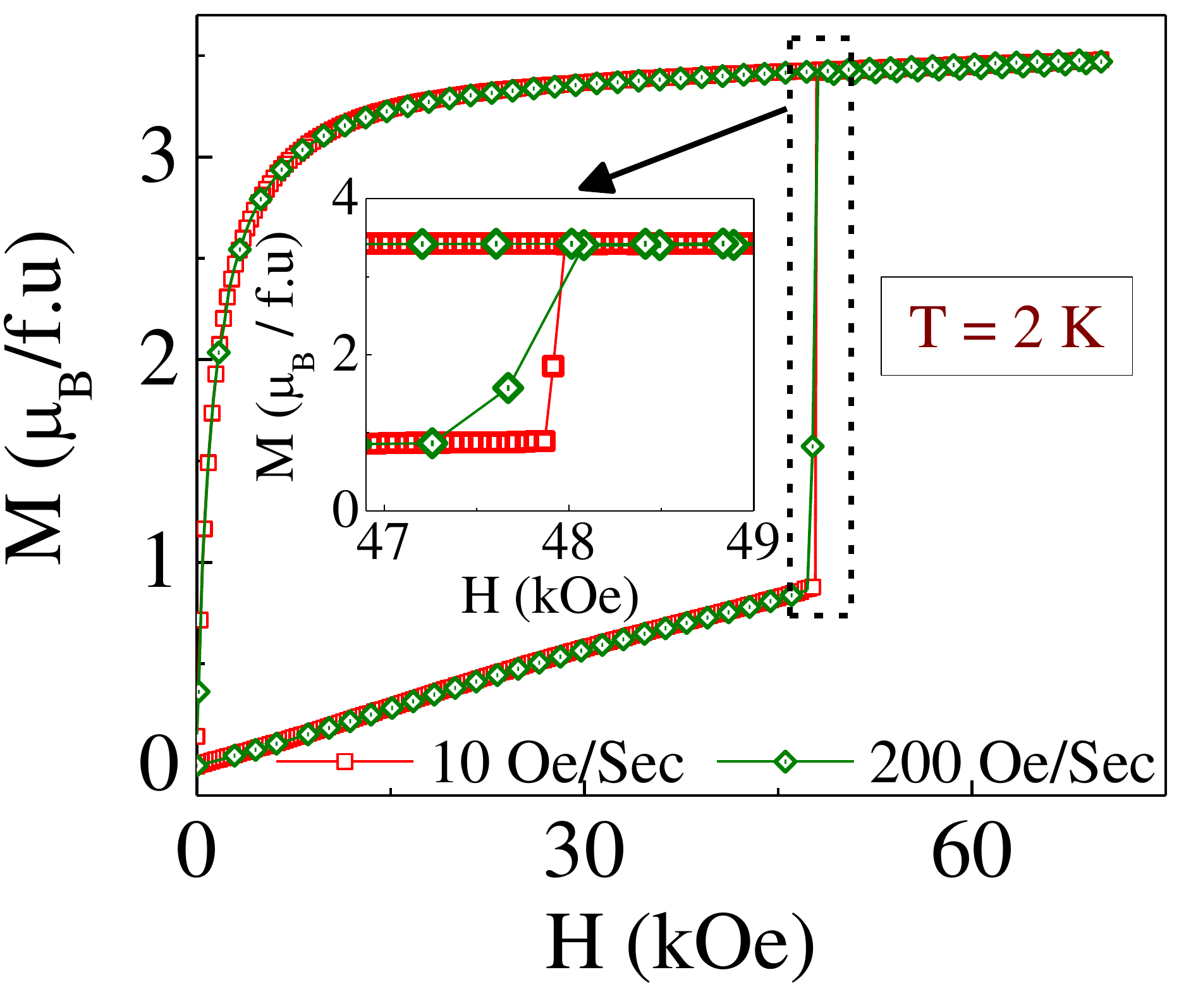}
\centering
\caption[]{Isothermal magnetization as a function of external magnetic field at 2 K temperature
for different field sweep rate 10 Oe/sec and 200 Oe/sec.}
\end{figure}
%**********************************************************************************

\subsection{Resistivity Relaxation study at 25 K}

To investigate the dynamics of the resistivity jumps we
performed resistivity relaxation measurements at T = 25 K (sample was ZFC from
room temperature down to 25 K) by applying three different magnetic fields
($H_{1},H_{2},H_{3}$) smaller than the $H_{CR}$ as shown in Fig. 11.
Resistivity of the sample as a function of time shows resistivity jumps at different
incubation time. The {\it incubation time}, time spent before the jumps, increases from
$\sim$20 Sec to $\sim$1000 Sec with decrease in the applied magnetic field. For the fixed temperature
(25 K), with decrease in the applied magnetic field the elastic barrier height (arises
due to the lattice strain at low temperature discussed earlier) increases and for that
reason system needs more {\it incubation time} to overcome the elastic barrier for
burst-like growth of the ferromagnetic fractions in the expense of the antiferromagnetic
components. This type of relaxation behavior is generally the characteristics of standard
martensitic transformations\cite{Maji,Lion,Wu,Shankaraiah,Hardy}.

%**********************************************************************************
\begin{figure}[h!]
\includegraphics[width=0.45\textwidth]{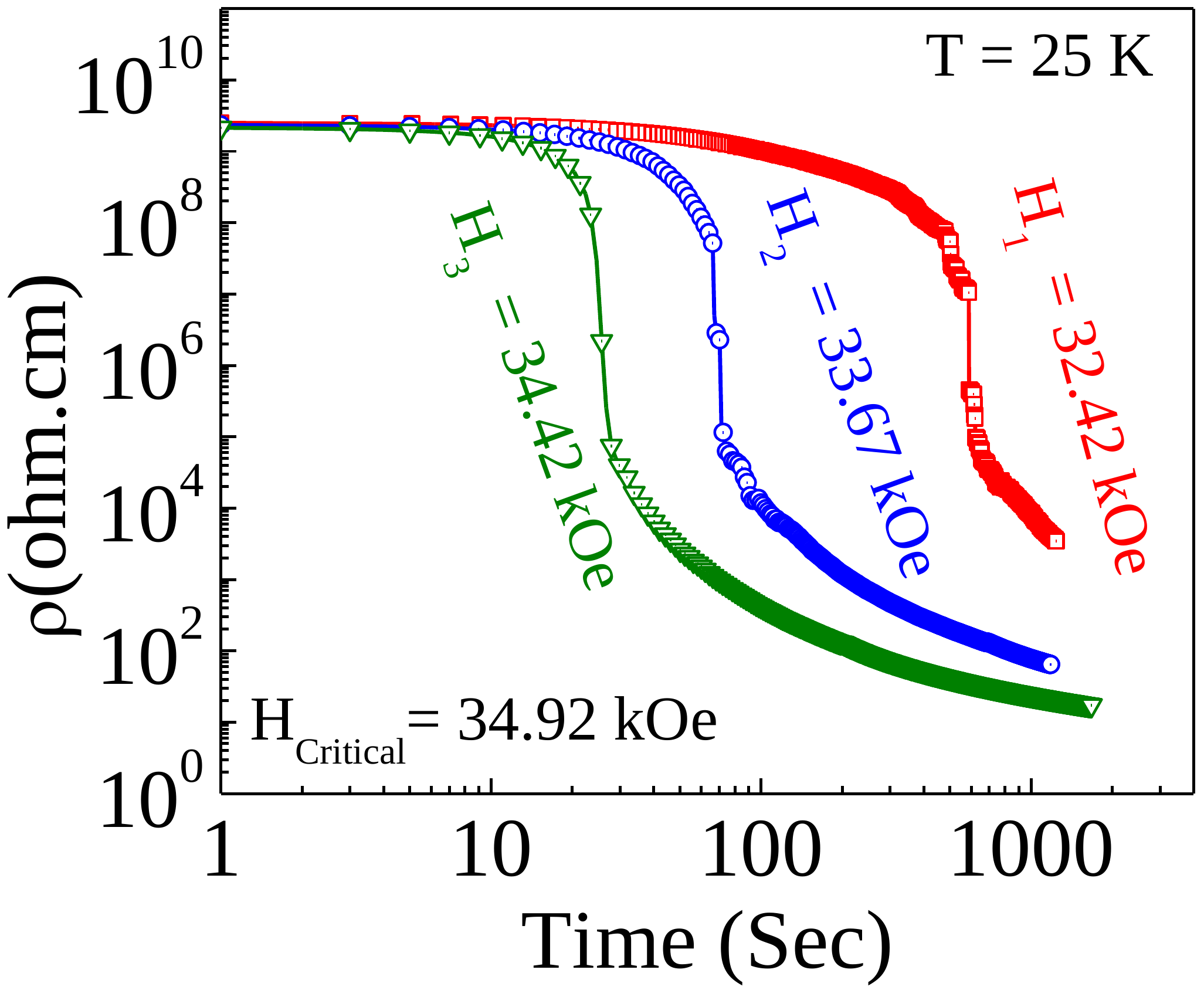}
\centering
\caption[]{Evolution of resistivity with time at 25 K for different
applied magnetic fields ($H_1,H_2,H_3$) such that ${H_1<H_2<H_3<H_{CR}}$.}
\end{figure}
%**********************************************************************************

\section*{II. Model Hamiltonian and Method}

We consider following two-band double-exchange model~\cite{dagotto1} for $e_g$
electrons, Hund's coupled to $t_{2g}$ (Mn core spins) in a square lattice:

\begin{eqnarray}
H &=& -\sum_{\langle ij \rangle \sigma}^{\alpha \beta}
t_{\alpha \beta}^{ij}
 c^{\dagger}_{i \alpha \sigma} c^{~}_{j \beta \sigma}
~ - J_H\sum_i {\bf S}_i.{\mbox {\boldmath $\sigma$}}_i \cr
&&
~~+ J_{AF}\sum_{\langle ij \rangle}
{\bf S}_i.{\bf S}_j
 - \lambda \sum_i {\bf Q}_i.{\mbox {\boldmath $\tau$}}_i
+ {K \over 2} \sum_i {\bf Q}_i^2 \cr
&&
+ {\sum_i(\epsilon_i - \mu) n_i}.
\end{eqnarray}

\noindent
Here, $c$ ($c^{\dagger}$) is the annihilation (creation) operator for $e_g$
electrons and $\alpha$, $\beta $ are the two Mn-$e_g$ orbitals $d_{x^2-y^2}$
and $d_{3z^2-r^2}$. ~$t_{\alpha \beta}^{ij}$~are
hopping amplitudes between nearest-neighbor sites in $x$ and $y$
directions:~$t_{a a}^x= t_{a a}^y \equiv t$,~$t_{b b}^x= t_{b b}^y
\equiv t/3 $,~$t_{a b}^x= t_{b a}^x \equiv -t/\sqrt{3} $,~$t_{a b}^y= t_{b a}^y
\equiv t/\sqrt{3}$.
$J_{\rm H}$ is the Hund's coupling between $e_{\rm g}$ electron spin
$\sigma_i$ and $t_{\rm 2g}$ spin
${\bf S}_{\rm i}$ at site $i$. The $e_{\rm g}$ electrons are also coupled to
Jahn-Teller phonons ${\bf Q}_{\rm i}$  by $\lambda$. $J$ is antiferromagnetic
super-exchange coupling between the $t_{\rm 2g}$ spins. We treat ${\bf S}_{\rm i}$
and ${\bf Q}_{\rm i}$ as classical variables~\cite{yunoki1,class-ref} and adopt
the double-exchange limit~\cite{dagotto1}
, i.e. $J_H \rightarrow \infty$. We set K (stiffness of Jahn-Teller modes) and
$|{\bf S}_{\rm i}|$ to be 1. This well studied model Hamiltonian qualitatively
reproduces the phase diagram of manganites.
The chemical potential $\mu$ is adjusted to fix the the electron density at
0.5. We include effect of disorder by adding $\sum_i \epsilon_i n_i$ term to
the Hamiltonian, where $\epsilon_i$ is the quenched disorder potential.
%We include the effect of disorder through an on site potential.

We applied Monte-Carlo simulation to the classical variables ${\bf S}_{\rm i}$ and
${\bf Q}_{\rm i}$, and an exact diagonalization scheme is employed to the fermionic
sector ($e_g$ electrons). We used travelling cluster approximation (TCA)~\cite{tca-ref}
based Monte-Carlo to handle large system size ($24 \times 24$ lattice). We annealed
the randomized classical spins ${\bf S}_i$ and lattice distortions ${\bf Q}_i$ in
an arbitrary quenched disorder configuration, and anneal down from temperature
$T$ = 0.1$t$ ($t$ is the hopping parameter).
The resistivity, in units of ${\hbar a}/{\pi  e^2 }$ ($a$: lattice constant)
is obtained by calculating the {\it dc} limit of the conductivity using the
Kubo-Greenwood formalism~\cite{mahan-book,cond-ref}. The magnetic structure factor
at wave vector ${\bf q} = (0, 0)$ is calculated from
$S(\textbf{q})$ = ${1 \over N^2}$ $\sum_{ij}$
$\bf {\bf S}_i\cdot {\bf S}_j$ e$^{i\bf{q} \cdot ({\bf r}_i-{\bf r}_j)}$.
Physical quantities, i.e. resistivity are averaged over ten different disorder
configurations in addition to the thermal averages taken during the Monte Carlo
simulations.

%------------------------------------------------------------------------------
%\section*{References}
%------------------------------------------------------------------------------

\end{document}